\documentclass[letterpaper,conference,10pt]{IEEEtran}

\usepackage{times}
\usepackage{graphicx}
\usepackage{graphics}
\usepackage{float}
\usepackage{amsthm}
\usepackage{amsmath}
\usepackage{cite}
\usepackage{xcolor}

\usepackage{array}
\usepackage{makecell}
\usepackage{booktabs}
\usepackage{bbold}
\usepackage[font=small]{caption}
\usepackage[font=small]{subcaption}
\usepackage{etoolbox}
\usepackage{threeparttable}
\usepackage[misc]{ifsym}

\usepackage{mathtools}
\usepackage[vlined,linesnumbered,ruled]{algorithm2e}
\SetAlFnt{\small\rmfamily}
\usepackage[hidelinks]{hyperref}
\usepackage[capitalize]{cleveref}
\renewcommand{\Cref}[1]{\cref{#1}}

\crefformat{footnote}{#2\textsuperscript{#1}#3}

\definecolor{darkgreen}{rgb}{0,0.6,0}

\AtBeginEnvironment{equation}{\small}
\AfterEndEnvironment{equation}{\normalsize}
\AtBeginEnvironment{align}{\small}
\AfterEndEnvironment{align}{\normalsize}
\AtBeginEnvironment{equation*}{\small}
\AfterEndEnvironment{equation*}{\normalsize}

\newtheorem{definition}{Definition}

\newcommand{\E}{\mathbb{E}}
\newcommand{\src}{\text{src}}
\newcommand{\tgt}{\text{tgt}}
\DeclareMathOperator{\dist}{dist}
\DeclareMathOperator{\iter}{iter}
\DeclareMathOperator{\precision}{precision@}
\newcommand{\preci}{{\precision}}
\newcommand{\round}[1]{\ensuremath{\lfloor#1\rceil}}

\SetKwFunction{Sample}{Sample}

\IEEEoverridecommandlockouts

\author{\IEEEauthorblockN{Han Zhang and Hong Xu\texorpdfstring{\textsuperscript{(\Letter)}}}\\
\IEEEauthorblockA{University of Southern California,
  Los Angeles, California, USA}\\
\texttt{zhan704@usc.edu, hongx@usc.edu}}

\begin{document}
\title{MANELA: A Multi-Agent Algorithm for Learning Network Embeddings}

\maketitle

\begin{abstract}
Playing an essential role in data mining, machine learning has a long history of being applied to networks on
multifarious tasks and has played an essential role in data mining. However, the discrete and sparse natures of networks often render it difficult to apply machine learning directly to networks. To circumvent this difficulty, one major school of thought to approach networks using machine learning is via \textit{network embeddings}. On the one hand, this network embeddings have achieved huge success on aggregated network data in recent years. On the other hand, learning network embeddings on \textit{distributively stored} networks still remained understudied: To the best of our knowledge, all existing algorithms for learning network embeddings have hitherto been exclusively \textit{centralized} and thus cannot be applied to these networks. To accommodate distributively stored networks, in this paper, we proposed a multi-agent model. Under this model, we developed the \textit{multi-agent network embedding learning algorithm} (MANELA) for learning network embeddings.  We demonstrate MANELA's advantages over other existing centralized network embedding learning algorithms both theoretically and experimentally. Finally, we further our understanding in MANELA via visualization and exploration of its relationship to DeepWalk.
\end{abstract}

\section{Introduction}

Playing an essential role in data mining, machine learning has a long history of being applied to networks on
multifarious tasks, such as network classification~\cite{snbgge08},
prediction of protein binding~\cite{adwf15}, etc. Thanks to the
advancement of technologies such as the Internet and database management
systems, the amount of data that are available for machine learning
algorithms have been growing tremendously over the past decade. Among
these datasets, a huge fraction can be modeled as networks, such as web
networks, brain networks, citation networks, street networks,
etc.~\cite{xskk18}. Therefore, improving machine learning algorithms on
networks has become even more important.

However, the discrete and sparse natures of networks often render it difficult to apply machine learning directly to networks. To circumvent this difficulty, one
major school of thought to approach networks using machine learning is via \textit{network
embeddings}~\cite{gf18}. A network embedding consists of a set of Euclidean
vectors, each of which represents a node in the network and encapsulates information about it. These vectors can
then be fed into machine learning algorithms as input for various classification and regression tasks.

Inspired by \textit{SkipGram}~\cite{msccd13} in the area of natural language processing, \cite{pas14} proposed \textit{DeepWalk}, a network embedding learning algorithm based on \textit{random walks} and neural networks. Theoretically, DeepWalk remains poorly understood. Nevertheless, thanks to its practical effectiveness, DeepWalk has become the core of most recently developed algorithms for learning
network embeddings, such as LINE~\cite{tqwzym15},
\textit{node2vec}~\cite{gl16}, DNGR~\cite{clx16}, metapath2vec~\cite{dcs17},
GraphSAGE~\cite{hyl17}, M-NMF~\cite{wcwpzy17}, LANE~\cite{hlh17},
PRUNE~\cite{lhcyl17}, RSDNE~\cite{wywwwl18}, and SIDE~\cite{kplk18}. These works are almost exclusively built upon random walks and neural networks.

Interestingly, to the best of our knowledge, all these works have been exclusively focusing on developing \textit{centralized} algorithms, a type of algorithm that critically relies on the computational resource of a single designated computer\footnote{This is commonly called a ``node'' in many other contexts. Here, we use the ``computer'' instead to avoid confusion with ``nodes'' as in a network.}, referred to as the \textit{central computer}. This, unfortunately, comes with at least two significant drawbacks.

\paragraph{It is often impossible to apply centralized algorithms to many networks that are distributively stored} Many common networks are inherently distributively stored. For example, consider a network of real-world email services. In such a network, nodes represent email accounts and edges represent correspondence between them. Nodes and edges are stored on different email servers and no single email server has access to the topology of the whole network. For another example, social networks have been perhaps the most commonly studied kind of networks in recent network embedding learning literature. In such a network, each node represents a social network account and each edge typically represents a friendship or follower/followee relationship between two accounts. In recent years, the growing concern over cyber privacy has led to the rise of \textit{federated social networks}, such as \texttt{Diaspora*}, \texttt{GNU Social}, and \texttt{pump.io}~\cite{srtmk17}. In a federated social network, each edge between two accounts are stored on the servers that host these two accounts and is normally inaccessible to other servers. Additional examples include business transaction/contracting networks, web of trust networks, etc.

Many of these networks share one common characteristic: Various considerations, such as legal concerns, privacy concerns, technical difficulties, etc. often erect an insurmountable barrier against scraping and aggregating them. This barrier, unfortunately, conflicts with the very foundation of centralized algorithms---That is, the ability to efficiently access all data from a single computer. Perhaps for this reason, these networks, despite their ubiquity, have been rarely studied in the data mining research community. 

To resolve this issue, it can be tempting to apply \textit{federated learning}~\cite{kmyrsb16}. While federated learning might be effective in many other contexts for learning distributively stored data, it fails to apply in the peculiar context of network embedding learning. In federated learning, a central computer coordinates other computers (e.g., users' mobile devices) with limited or no sharing of data between them. However, at least one computer inevitably has access to all learning results, i.e., the whole learned network embedding. Since each vector in the network embedding encapsulates information about a node, federated learning effectively aggregates information about the network and therefore still leaves the issue unresolved.

One possible direction to address this issue is to develop \textit{multi-agent} algorithms. In the context of network embeddings, an agent can be seen as an autonomous unit that is circumscribed to have access to only a local subset of the network, such as an email server or a federated social network server in the aforementioned examples. An algorithm is multi-agent if agents are able to cooperatively carry it out with communications that do not disclose excessive information to other agents. This locality nature makes a multi-agent algorithm well-suited for learning embeddings on these networks.
 
\paragraph{Even for networks that are aggregated, centralized algorithms still bear scalability issues} As the amount of the input data and the complexity of the model increase, such algorithms would inevitably require \textit{vertical scaling}, i.e., adding resources (such as RAM, CPUs, GPUs, etc.) to the central computer. Vertical scaling, however, usually comes with serious pecuniary limitations: It becomes extremely expensive once the required amount of resources has increased beyond a certain point~\cite{fa11}. Unfortunately, this is undesirable and can be even devastating: It prohibits against fully utilizing and building models with complexities commensurate with the massive amount of available data in the era of ``big data,'' where networks with over billions of nodes are commonplace.

Facing the scalability issue of centralized algorithms, many researchers have turned to \textit{distributed} algorithms. Unlike centralized algorithms, which always critically rely on a central computer, distributed algorithms are capable of taking advantages of distributed computational resources. They can be also seen as generalizations of multi-agent algorithms where the circumscription on information accessibility is lifted. Instead of relying on vertical scaling, a distributed algorithm connotes \textit{horizontal scaling}: More computational resources can be added with only a linearly increasing cost by simply incorporating more computers. In fact, in many areas of machine learning and data mining, researchers have started paying more attention to distributed algorithms. For instance, in the area of deep learning, a huge portion of the TensorFlow project has been focusing on developing algorithms that train deep neural networks under various distributed computational environments~\cite{tensorflow}. Since a multi-agent algorithm is always distributed, an efficient multi-agent algorithm may also potentially ameliorate the scalability issue.

Because of the importance of network embedding learning and its aforementioned contemporary
challenges, i.e., inherently distributively stored networks and the scalability issue, it has become imperative to develop multi-agent algorithms for learning network embeddings.
Surprisingly, to the best of our knowledge, until today, there do not yet exist any multi-agent algorithms for learning network embeddings.

\textbf{Our contributions:} In this paper, we develop the first multi-agent algorithm for learning network embeddings.
We present our new multi-agent model (\cref{sec:multi-agent}) and our new multi-agent algorithm (\cref{sec:algorithm}), the \textit{multi-agent network embedding learning algorithm} (MANELA), for learning network embeddings under our multi-agent model. We also compare MANELA with other baseline algorithms both theoretically (\cref{sec:theory}) and experimentally (\cref{sec:experiment}). Finally, we further our understanding of MANELA via visualization (\cref{sec:visualization}) and exploration of its relationship to DeepWalk (\cref{sec:relation-to-deepwalk}).

\section{Background}\label{sec:background}
In this section, we introduce background on network embeddings, random walk-based algorithms, and multi-agent algorithms.

\subsection{Definitions and Notations}
\Cref{tab:notations} shows a list of basic notations and their meanings.
\begin{table}[t]
    \centering
    \caption{Notations used in this paper.}\label{tab:notations}
    \begin{tabular}{ c l }
    \toprule
        \textbf{Notation} & \textbf{Definition} \\
        \midrule
        \(G\) & a network\\
        \(V\) & the node set of \(G\)\\
        \(E\) & the edge set of \(G\)\\
        \(|\cdot|\) & the number of elements in \(\cdot\)\\
        \(v,u\) & nodes in \(V\) \\
        \(\mathcal{N}(v)\) & the set of all neighbors of \(v\)\\
        \(a(v)\) & the agent that maintains \(v\)\\
        \(d(v)\) & the degree of \(v\)\\
        \(\dist(v,v')\) & the distance between \(v\) and \(v'\) on \(G\)\\
        \(\Phi(v)\) & an Euclidean vector that represents \(v\)\\
        \(f(v)\) & the frequency at which \(a(v)\) performs updates\\
     
        \bottomrule
    \end{tabular}
\end{table}
Given a network \(G=\langle V, E \rangle\), its embedding is a set \(\{\Phi(v)\mid v\in V\}\) of \(m\)-D Euclidean vectors, in which each vector \(\Phi(v)\) represents a node \(v\in V\). These vectors encode some semantics of these nodes, such as their adjacency and structural information in \(G\). 

\subsection{Random Walk-Based Algorithms}

Random walk-based algorithms are a family of algorithms that learn network embeddings from paths sampled using random walks in the input network. We refer to such paths as \textit{walk paths}. Generally speaking, these algorithms learn the embedding of each node in the network by minimizing distances (under a certain kind of metric) between the vectors that represent nodes co-occurring in walk paths. They usually perform these minimizations via neural networks of simple architectures. We introduce two representative algorithms in detail:

\subsubsection{DeepWalk}
DeepWalk~\cite{pas14} undoubtedly plays an essential role in the network embedding learning algorithm development in the past few years. Most, if not all, random walk-based algorithms after the proposal of DeepWalk were more or less variants of it. Inspired by \textit{SkipGram}~\cite{msccd13} in the area of natural language processing, DeepWalk applies a simple neural network on walk paths. It first generates multiple walk paths of length \(\ell\) starting at each node for \(\gamma\) times, where \(\ell\) and \(\gamma\) are hyperparameters.
Then it treats nodes in each walk path \((v_1,\ldots,v_{\ell})\) as if they were words in each sentence (as in SkipGram)---that is, it maximizes the probability of the presences of the nodes within distance \(k\) to any node \(v_{i}\) given the presence of \(v_{i}\) in a walk path, i.e., maximizing \(\sum_{i=1}^{\ell} \log \Pr(v_{\max\{1,i-w\}},\allowbreak \ldots,v_{i-1},\allowbreak v_{i+1},\allowbreak\ldots,\allowbreak v_{\min\{\ell,i+w\}}\mid v_{i})\), where \(w\) is a hyperparameter, referred to as the \textit{window size}. Either \textit{hierarchical softmax} or \textit{negative sampling} can be used to approach this optimization problem.
\subsubsection{node2vec}
Node2vec~\cite{gl16} is a variant of DeepWalk that replaces its simple random walk strategy with a more sophisticated alternative. Node2vec provides hyperparameters \(p,q\) to adjust how fast its biased random walks would move further away from their predecessing nodes in a walk path. \(p\) and \(q\) are competitor hyperparameters such that larger \(p\) results in further walks from the predecessing nodes, whereas larger \(q\) renders closer walks. This biased random walk strategy is a middle ground between breadth-first search (BFS) and depth-first search (DFS) to strike a balance between encapsulating different causes of similarities between nodes.

\subsection{Multi-Agent Algorithms}
A multi-agent algorithm is an algorithm that is carried out by agents via local computation and circumscribed communication. Agents are autonomous units that have access to local information, e.g., information about itself and its locally accessible neighbors. A multi-agent algorithm usually permits communication between agents locally and without disclosing excessive information, such as those acquired from another agent that is neither of the two communicating parties.

We note that, another distinct concept, a distributed algorithm, is often confused with a multi-agent algorithm. In contradistinction to a centralized algorithm, a distributed algorithm is simply a kind of algorithm that is capable of relatively more efficiently taking advantage of multiple computers. While, by nature, a multi-agent algorithm is always a distributed algorithm, the demarcation between a multi-agent algorithm and a general distributed algorithm lies in the agents' autonomy: For example, a distributed algorithm would enjoy the benefits brought by a centralized computer and full communication between computers, which are, however, unacceptable for a multi-agent algorithm.

\section{Our Multi-Agent Model}\label{sec:multi-agent}

The exact formulation of ``multi-agent'' varies from application to application. In the context of network embedding learning, we formalize the multi-agent model \(\langle G=\langle V, E\rangle, s \rangle\) as follows:
\begin{enumerate}
    \item \textbf{Maintenance}: On the given network \(G\), each node \(v\in V\) is maintained by an agent \(a(v)\).
    \item \textbf{Communication}: Every agent knows the existence of all other agents and is permitted to communicate with any of them.
    \item \textbf{Possession of \(G\)}: Each agent \(a(v)\) by itself only directly possesses knowledge on all edges incident to \(v\).
    \item \textbf{Accessibility of \(G\)}: The accessibility of \(G\) to \(a(v)\) is circumscribed: Each agent \(a(v)\) is permitted and is only permitted to know the subnetwork induced by all nodes within a small distance \(s\) to \(v\)---In other words, each agent is permitted to know a defined portion of \(G\); it can acquire knowledge on \(G\) via communication with other agents if it does not possess this knowledge but is permitted to know; but it is prohibited from acquiring knowledge on \(G\) beyond its permission by any means including communication with (chains of) other agents.
\end{enumerate}
We refer to these restrictions as the \textit{accessibility circumscription} (AC). The task is to learn a vector representation for each node under the AC\@.

Our multi-agent model is applicable to a variety of real-world scenarios. For example, let us consider a Facebook-style federated social network such as \texttt{Diaspora*}. Each user can be seen as a node \(v\) in \(G\) and two nodes are connected iff their corresponding users are friends. (1): \(a(v)\) is the server that hosts \(v\). (2): There is a global user directory so that each user knows the existence of all other users and is permitted to communicate with any of them. (3) and (4): \(a(v)\) possesses \(v\)'s friend list (adjacent nodes) is permitted to inquire \(v'\)'s host \(a(v')\) about \(v'\)'s friend list if \(v\) and \(v'\) are adjacent to each other (and hence \(s=2\)), but is prohibited (or undesirable due to technical reasons) from storing them. For privacy concerns, \(a(v)\) is also prohibited from inquiring \(v''\)'s friend list if \(v\) and \(v''\) are not adjacent to each other.

\section{Our Algorithm: MANELA}\label{sec:algorithm}

In this section, we first discuss the inadaptability of random walk-based algorithms to be multi-agent. Then we outline the framework of our multi-agent algorithm for learning network embeddings. Thenceforth, we delineate this framework with discussions.

\subsection{Inadaptability of Random Walk-Based Algorithms}

The well known centralized network embedding learning algorithm, DeepWalk, along with its variants, has achieved far better effectiveness than their traditional non-embedding-based counterparts on many tasks~\cite{pas14,gl16}. To design a multi-agent algorithm, the most obvious means is perhaps to derive one from these algorithms. Unfortunately, the centralized nature of random walks and the agents' autonomy in a multi-agent model can hardly reconcile: Walk paths inherently bury excessive information about the network. For example, the frequency at which a node \(v\) is present in walk paths is accessible to \(a_v\) but also embodies the information about the distribution of degrees of nodes in the whole network. This irreconciliation, referred to as the \textit{AC hazard}, undermines the very foundation of multi-agent algorithms, i.e., the circumscription on information accessibility to agents. 

\subsection{Outline of Our Multi-Agent Algorithm}

Facing the AC hazard, it is imperative to develop a new algorithmic framework that accommodates multi-agent network embedding learning algorithms. Observing that random walk, in essence, is a method for sampling pairs of nodes as training data in random walk-based algorithms, we only need to substitute it with a sampling method that comports with the requirement of our multi-agent model.

Similar to Equation~(4) in~\cite{msccd13}, for a given network \(G=\langle V, E\rangle \), we model our maximization objective as
\begin{equation}
\begin{split}\label{eq:manela-opt}
    &\sum_{\substack{v\in V}}\bigg\{ \sum_{i=1}^{2w} \E_{v'_i\sim M\left(v,s,\vec{r}\right)} \log \sigma\left (\Phi(v)^{\top} \Phi(v'_i)\right) +\\
    &\sum_{i=1}^{2w}\sum_{j=1}^k \E_{v'_{ij}\sim P_G}\left[\log \sigma \left(-\Phi(v)^{\top} \Phi(v'_{ij})\right ) \right] \bigg\}.
    \end{split}
\end{equation}
Here, \(v'_i\) is subject to the distribution \(M\left(v,s,\vec{r}\right)\), which we will discuss in detail later in \cref{sec:m}.
\(w\), called the \textit{half sample size}, is a scalar hyperparameter that is analogous to DeepWalk's window size. \(\vec{r}\), called the \textit{ratio vector}, is a normalized \(s\)-dimension vector hyperparameter in which the sum of all components is one. \(v'_{ij}\) is subject to distribution \(P_G\) in a negative sampling procedure, which we choose as uniform distribution in our algorithm. 

Our algorithm is a multi-agent algorithm in which the vector \(\Phi(v)\) representing each node \(v \in V\) is maintained by agent \(a(v)\) and is only directly accessible to it. We refer to our algorithm as MANELA\@. To explain MANELA, we first define two subprocedures: \textit{agent \(a(v)\)'s single positive/negative update on \(v\) from node \(v'\)}. During \(a(v)\)'s single positive update on \(v\) from \(v'\), \(a(v)\) requests for \(\Phi(v')\) from \(a(v')\). Then, \(a(v)\) updates \(\Phi(v)\) using
\begin{equation}
\begin{split}\label{eq:manela-update}
\Phi(v)\gets &\Phi(v) + \alpha \cdot \frac{\partial \log \sigma\left (\Phi(v)^{\top} \Phi(v')\right )}{\partial \Phi(v)}\\
=&\Phi(v) + \alpha \cdot \left(1 - \sigma \left (\Phi(v)^{\top} \Phi(v')\right )\right)\cdot \Phi(v'),
\end{split}
\end{equation}
where \(\alpha\) is the learning rate. \(a(v)\)'s single negative update on \(v\) from \(a(v')\) is similar to its positive counterpart except that \(\Phi(v')\) is replaced with \(-\Phi(v')\).

\begin{algorithm}[t]
    \DontPrintSemicolon
    \SetKwRepeat{Repeat}{repeat}{until}
    \SetKwFor{AtFreq}{at frequency}{do}{end}
    \KwIn{\(v\): the node that the agent maintains}
    \KwIn{\(m\): the dimension of the vector that represents \(v\)}
    \KwIn{\(w\): the half sample size}
    \KwIn{\(\vec{r}\): the ratio vector}
    \KwIn{\(\kappa\): number of negative samples}
    \KwIn{\(T\): maximum running time}
    \KwOut{the vector that represents \(v\)}
    Initialize \(\Phi(v)\) to a \(m\)-D vector of random numbers\;\label{alg:initialization}
    \Repeat{current running time exceeds \(T\)\label{alg:time-limit}}{
    \AtFreq{\(f(v)\)\label{alg:freq}}{
      \ForEach{\(v_+\in \Sample(v,w,\vec{r})\)}{\label{alg:positive-sample}
        \(\Phi(v) \gets\) Single positive update on \(v\) from \(v_+\)\;\label{alg:positive-update}
        Uniformly randomly sample \(\kappa\) nodes from \(V\) into a newly created set \(V_-\)\;\label{alg:negative-sample}
        \ForEach{\(v_-\in V_-\)}{\label{alg:negative-update-begin}
          \(\Phi(v) \gets\) Single negative update on \(v\) from \(v_-\)}\;\label{alg:negative-update-end}
        }
      }
    }
    \Return \(\Phi(v)\)\;
    \caption{Update algorithm of agent \(a(v)\)}\label{algo:agent}
\end{algorithm}

\Cref{algo:agent} outlines MANELA and it works as follows. Each agent \(v\) initializes \(\Phi(v)\) to a random \(m\)-D vector (\cref{alg:initialization}). For each node \(v\in V\), at a certain frequency \(f(v)\) (\cref{alg:freq}), agent \(a(v)\) samples \(v_+\) from distribution \(M\left(v,s,\vec{r}\right)\) (\cref{alg:positive-sample}) and performs its single positive update on \(v\) from \(a(v_+)\) once (\cref{alg:positive-update}). In between \(a(v)\)'s every two consecutive single positive updates, \(a(v)\) randomly samples \(\kappa\) nodes \(v_{-,1},\ldots v_{-,\kappa}\) (\cref{alg:negative-sample}) and performs its single negative updates on \(v\) from each of them (\crefrange{alg:negative-update-begin}{alg:negative-update-end}), which corresponds to a \(P_G\) that is uniform over \(V\). MANELA terminates upon reaching a running time limit (\cref{alg:time-limit}).

\subsection{Discussion: Iteration Frequency}
What is the best frequency for a specific agent to perform updates? For terminological convenience, we first make the following definitions. 
\begin{definition}
An \textit{update pair} \(\langle v_{\src},v_{\tgt}\rangle\), refers to the updates that consist of \(a(v_{\src})\)'s single positive update on node \(v_{\src}\) from node \(v_{\tgt}\) and its subsequent \(\kappa\) single negative updates from randomly sampled nodes on node \(v_{\src}\) (\crefrange{alg:positive-sample}{alg:negative-update-end} in \cref{algo:agent}). Here, \(v_{\src}\) is referred to as the \textit{source node} and \(v_{\tgt}\) is referred to as the \textit{target node}.
\end{definition}

The concept of update pair can be generalized to DeepWalk and node2vec if negative sampling is used, where we refer to an update pair as a positive update between \(v_{\src}\) and \(v_{\tgt}\) and the subsequent \(\kappa\) negative updates between \(v_{\src}\) and other nodes.

\begin{definition}
An \textit{iteration} \(\iter(v_{\src})\) is a set of source node related streaming update pairs performing in a short time:
\begin{equation}\label{eq:iter_def}
    \iter(v_{\src})\coloneqq \{<v_{\src},v_{\tgt}^i> \mid i=1,\ldots,2w,v_{\tgt}^i\sim M(v_{\src},s,\vec{r})\}.
\end{equation}
\end{definition}

We define the \textit{frequency hyperparameter}, denoted by \(f(v)\), as the number of iterations for which agent \(a(v)\) performs in unit time. A policy that naively specifies a uniform frequency hyperparameter for all agents would be unviable. As an illustration, let us suppose node \(v\) has thousands of adjacent nodes, while node \(v'\) has only a few. Intuitively, \(a(v)\) is supposed to perform more updates than \(a(v')\) because \(v\)'s local information in the network is far richer than \(v'\) and deserves more computing resources. Therefore, in our algorithm, we set the frequency at which an agent \(a(v)\) performs updates to be proportional to \(v\)'s degree \(d(v)\), i.e., \(f(v) \propto d(v)\).

\subsection{Discussion: \(M(v,s,\vec{r})\) }\label{sec:m}
We now discuss the distribution \(M(v,s,\vec{r})\), from which the target nodes in update pairs are sampled. In a network, there are two widely accepted \textit{causes that lead to the similarity between nodes} (CSNs): \textit{homophily} and \textit{structural equivalence}. Homophily refers to that, the smaller the distance between two nodes, the more similar they are. For example, in a social network, there are coteries of directly connected people (i.e., nodes) who share common interests and they are likely to be similar to each other. Structural equivalence refers to that the similarity of two nodes are commensurate with that of their structural roles in their respective vicinities~\cite{gf18}. For example, once again in a social network, there are fans (i.e., nodes) who are connected to the same celebrities (i.e., nodes) and these fans are likely to be similar to each other.

\cite{gl16} observed that random walks guided by BFS result in embeddings that encapsulate more structural equivalence than homophily, while random walks guided by DFS result in embeddings that encapsulate more homophily than structural equivalence. In light of this observation, we choose a middle ground between the two sampling strategies to trade off the encapsulation of homophily and structural equivalence in our algorithm. To achieve this, we embody the trade-off into the distribution \(M\), such that altering its hyperparameters leads to different trade-offs between homophily and structural equivalence. Formally, we define \(M(v,s,\vec{r})\) as satisfying
\begin{align}
       P\left(\dist(v',v)=k; \vec{r}, s\right)&=\begin{cases}r_k,\quad &1\le k \le s\\0,\quad &\text{otherwise}\end{cases}\label{eq:m-condition}\\
       P\left(v'\mid \dist(v',v)=k; v\right) &\propto 1,\label{eq:m-prob}
\end{align}
where \(r_k\) is the the \(k^{\text{th}}\) component of hyperparameter \(\mathbf{r}\). In \cref{eq:m-condition}, \(M(v,s,\vec{r})\) distinguishes nodes in the network based on their distances to the source node \(v\): The probability that the distance from the target node \(v'\) to \(v\) is \(k\) is specified by \(r_k\). In \cref{eq:m-prob}, the conditional distribution when the distance is \(k\) is a uniform distribution. \(\vec{r}\) is a vector hyperparameter that can be used to adjust the sampling weights of nodes at different distances from \(v\). At one extreme, if \(r_1=1\) and all other components are set to 0, then obviously all sampled target nodes are adjacent to \(v\). As a result, the sampled nodes are similar to those sampled by random walks guided by BFS, and thus the learned network embeddings would encapsulate more structural equivalence than homophily. At the other extreme, if we set \(r_1=r_2=\ldots =r_s\), then the sampled nodes are similar to those sampled by random walks guided by DFS, and thus the learned network embeddings would encapsulate more homophily than structural equivalence.

In practice, such distributions are difficult to sample: Even if \(s\) is as small as 2, it may demand an exorbitant amount of computational resources for an agent \(a(v)\) to store \(v\)'s distances from all nodes within distance \(s\).
Therefore, we introduce an approximation sampling strategy based on the following definition:
\begin{definition}
    A node \(v'\) is a \textit{\(k\)-hop neighbor} to \(v\) on \(G\) if there exists a path of length \(k\) on \(G\) that starts at \(v\) and ends at \(v'\).
\end{definition}
Our sampling strategy is that, in each iteration, for each \(k\in [1,s]\), agent \(a(v)\) uniformly samples \(\round{2wr_{k}}\) from all \(v\)'s \(k\)-hop neighbors as target nodes. The size of the samples is approximately \(2w\). \Cref{algo:Sample} shows the details of our sampling strategy.

\begin{algorithm}[t]
    \DontPrintSemicolon
    \SetKwFor{RepeatTimes}{repeat}{times}{end}
    \SetKwFunction{Merge}{Merge}
    \KwIn{\(v_{\src}\): the source node}
    \KwIn{\(w\): the half sample size}
    \KwIn{\(\vec{r}\): the ratio vector}
    \KwOut{a list of sampled nodes}
    Initialize \(S\) to an empty list\;
    \For{\(k\gets 1\) \KwTo \(|\vec r|\)}{
    \RepeatTimes{\(\round{2 w\times r_k} \)}{
        \(v\gets v_{\src}\)\;
        \RepeatTimes{k}{
        Uniformly randomly sample a node \(v'\) from \(\mathcal{N}(v)\)\;
        \(v\gets v'\)\;
        }
        Add \(v\) to \(S\)\;
    }
    }
    \Return \(S\)\;
    \caption{Function: Sample(\(v_{\src}, w, \vec{r}\))}\label{algo:Sample}
\end{algorithm}

\section{Theoretical Analysis}\label{sec:theory}

\begin{table*}[t]
    \centering
    \caption{Theoretical comparison between different algorithms. Subscripts ``sa,'' ``up,'' and ``pre'' mean the sampling, update, and preprocessing subprocedures, respectively. In the third column, ``both'' indicates that both homophily and structural equivalence can be encapsulated; and ``tunable/untunable'' indicates whether the encapsulation of different CSNs can be tuned.}
    \label{tab:comparison}
    \begin{threeparttable}
    \centering
    \begin{tabular}{ccccc}
    \toprule
        \textbf{Algorithm} &\phantom{a}&\textbf{Amortized Time Complexity}&\textbf{Capability of Encapsulating Different CSNs } & \textbf{Multi-Agent?}\\
    \midrule
        MANELA && \(\Theta_{\text{sa}}(N_{\text{u}}(G))\tnote{$\ddag$}~+\Theta_{\text{up}}\left( N_{\text{u}}(G) \right) \) & Both, Tunable & Yes \\
        DeepWalk && \(\Theta_{\text{sa}}\left(N_{\text{u}}(G)\right)\tnote{*}~ + \Theta_{\text{up}}\left( N_{\text{u}}(G) \right)\)& Both, Untunable&No\\
        node2vec && \( O_{\text{pre}}\left(|E|\cdot |V|\right) + \Theta_{\text{sa}}\left(N_{\text{u}}(G)\right)\tnote{*}~ + \Theta_{\text{up}}\left( N_{\text{u}}(G) \right)\) & Both, Tunable & No\\

    \bottomrule
    \end{tabular}
    \begin{tablenotes}[para]\footnotesize
    \item[$\ddag$] \(\Theta_{\text{sa}}(s\cdot N_{\text{u}}(G))\) if \(s\) is not treated as a constant.
    \item[*] \(\Theta_{\text{sa}}\left(\frac{\ell}{w(2\ell-w-1)}\cdot N_{\text{u}}(G)\right)\) if \(\ell\) and \(w\) are not treated as constants.
    \end{tablenotes}
    \end{threeparttable}
\end{table*}

In this section, we theoretically compare MANELA, DeepWalk, and node2vec in terms of time complexity and capability of encapsulating different CSNs. Throughout this section, for the sake of simplicity in our theoretical analysis, we assume that the input network is always connected. To the best of our knowledge, this is also the first theoretical analysis of the time complexity and the capability of encapsulating different CSNs on DeepWalk and node2vec.

\subsection{Time Complexity}

In this subsection, we analyze the time complexities of the network embedding learning algorithms. We analyze MANELA, DeepWalk, and Node2vec in three parts: the sampling subprocedure, and the update subprocedure, and any preprocessing subprocedure (node2vec only). For the sake of fair comparison, we assume that DeepWalk and node2vec use negative sampling in lieu of hierarchical softmax (which is generally more costly). We assume that all algorithms are required to perform no less than the same number \(N_{\text{u}}(G)\) of times of update pairs for each node in the network as the source node and share a same constant \(\kappa\), where \(N_{\text{u}}(G)\) is a monotonically non-decreasing function with respect to \(|V|\) and \(|E|\). Practically speaking, we would desire \(N_{\text{u}}=\Omega(|V|)\) for a reasonable effectiveness and \(N_{\text{u}}(G)=O(|E|)\) for a reasonable scalability.

\subsubsection{The Sampling Subprocedure}
For random walk-based algorithms, a walk path samples \(\ell\) target nodes that are used for multiple source nodes. In such a walk path, \(2w\) nodes are used as target nodes for each of the middle \((\ell-2w)\) nodes, whereas all the rest \(2w\) nodes near either end of the path use a total of \(2(w+(w+1)+\ldots+(2w-1))=w(3w-1)\) nodes as target nodes. The number of performed update pairs with the source node being in the walk path is thus \(2w(\ell-2w)+w(3w-1)=w(2\ell-w-1)\). The resulted amortized time complexity of the sampling subprocedure is \(\ell / w(2\ell-w-1)\) for each update pair. Therefore, the sampling subprocedures in DeepWalk and node2vec have a time complexity of \(\Theta((\ell / w(2\ell-w-1))\cdot N_\text{u}(G))\). Practically, \(\ell\) and \(w\) are usually fixed as constants. In this case, the time complexity is simplified to \(\Theta(N_{\text{u}}(G))\).

In MANELA, the amortized time complexity of the sampling subprocedure for each update pair is \(O(s)\). Therefore, MANELA's sampling subprocedure has a time complexity of \(\Theta(s\cdot N_{\text{u}}(G))\). Practically, \(s\) is usually a very small constant and does not change with respect to \(G\). In this case, this time complexity is simplified to \(\Theta(N_{\text{u}}(G))\).

\subsubsection{The Update Subprocedure}
All three embedding learning algorithms share the same time complexity of \(O(1)\) per update pair. The total time complexity for the update subprocedure is thus \(\Theta(N_{\text{u}}(G))\).

\subsubsection{The Preprocessing Subprocedure}\label{sec:node2vec-pre}
Node2vec is the only algorithm in our analysis that consists of a preprocessing subprocedure for preparing a probability matrix that guides its random walks. For each \(v\in V\) and all pairs of its adjacent nodes \(v_i\) and \(v_j\), it computes the matrix \(\{p_{ij}=P\left(suc=v_{j}\mid pre=v_{i}\right)\}\). \(p_{ij}\) indicates the probability that \(v_j\) is the successor of \(v\) under the condition that \(v_i\) is the predecessor of \(v\) in a walk path. Therefore, the time complexity of this preprocessing subprocedure is \(\Theta\left (\sum_{v\in V}d(v)^2\right )\), and we have
\begin{align}
    \sum_{v\in V}d(v)^2 &\geq \frac{(\sum d(v))^2}{|V|}=\frac{4|E|^2}{|V|}\label{eq:n2v-omega}\\
    \sum_{v\in V}d(v)^2 &= O\left({|V|}^2\cdot \frac{|E|}{|V|}+\left(|V|-\frac{|E|}{|V|}\right)\right)=O\left( |E|\cdot |V|\right ).\label{eq:n2v-o}
\end{align}
Here, \cref{eq:n2v-omega} applies the Cauchy-Schwarz inequality and hence proves that this preprocessing step has a time complexity of \(\Omega (|E|^2/|V|)\). \cref{eq:n2v-o} considers the worst scenario where edges are allocated in a way so as to maximize the number of nodes that are connected to all other nodes. There can be no more than \(O(|E|/|V|)\) nodes that are connected to all other nodes and their degrees are \(O(|V|)\). With all other nodes having a degree of \(O(1)\), we have the term to the right of the first equal sign. Therefore, \cref{eq:n2v-o} proves that this preprocessing step also has a time complexity of \(O(|E|\cdot |V|)\). In addition, this upper bound is tight: Let us consider a complete network. Then we have \(\sum_{v\in V}d(v)^2=|V|\cdot {\left(|V|-1\right)}^2= O(|E|\cdot |V|)\), which reaches the upper bound.

On the one hand, the worst-case running time increases at least quadratic with respect to the number of nodes. On the other hand, the best-case running time also increases superlinearly insofar as the network is not sparse.\footnote{Our theoretical analysis on node2vec's scalability here seemingly contradicts \cite[Figure~6]{gl16}, which empirically demonstrates that node2vec's running time increased linearly with respect to the number of nodes on a series of Erd{\H o}s-R{\' e}nyi networks. We can reconcile these two results as follows. With a fixed average degree of nodes, we expect \(|E|=\Theta(|V|)\) for Erd{\H o}s-R{\' e}nyi networks and therefore they are sparse. Then they might have just happened to fall in node2vec's best-case scenarios and hence demonstrated an empirically linear scalability.} We do not think this is unscalable and, therefore, it is unlikely that node2vec can survive today's networks that commonly have more than millions of nodes.

\subsection{Capability of Encapuslating CSNs}
We now turn to their capability of encapsulating the two kinds of CSNs: homophily and structurual equivalence. All three algorithms are capable of encapsulating both of them with some trade-off. DeepWalk provides no hyperparameter for tuning this trade-off. Unlike DeepWalk, node2vec enables such tuning, by providing hyperparameters for adjusting how fast its biased random walks would move further away from their predecessing nodes in walk paths. MANELA enables such tuning via \(\vec r\).

\subsection{Summary}

We have summarized our results in \cref{tab:comparison}. From all three perspectives listed in the table, MANELA always has the best results. Practically speaking, MANELA's time complexity is the same as that of DeepWalk and is much smaller than that of nod2vec, which we have proven to have a serious scalability issue. The encapsulation of different CSNs is tunable in MANELA and node2vec, but not in DeepWalk. Most importantly, neither DeepWalk nor node2vec can be made multi-agent, but MANELA is multi-agent and is thus more widely applicable.

\section{Experimental Evaluation}\label{sec:experiment}

In this section, we experimentally evaluate MANELA on two tasks: node classification and link prediction.
We ran our experiments in Windows Server R2 on an Amazon Elastic Compute Cloud (Amazon EC2) instance with 32GB RAM\@. For DeepWalk and node2vec, we used the implementations provided by their respective authors. We implemented all other algorithms in Python 3.

\subsection{Implementation of MANELA}
We followed the algorithmic details in \cref{sec:algorithm} and implemented MANELA under centralized settings (for evaluational purposes)\footnote{\url{https://github.com/hanzh015/MANELA}}.
To eliminate undesirable uncertainties and thus to promote fairness in experimental comparisons, instead of setting a maximum running time \(T\), we set a maximum number of update pairs. The underlying rationale is that, once \(\kappa\) and \(w\) are fixed, each update pair would take approximately equal amount of time. We also fix \(s=2\) throughout \cref{sec:experiment,sec:visualization} due to its broad real-world applicability. As a result, \(\vec{r}\) is 2-D and hence \(r_2\) can be determined using \(r_2=1-r_1\) once \(r_1\) is known. For this reason, we use \(r_1\) in lieu of \(\vec r\) when describing experimental settings throughout \cref{sec:experiment,sec:visualization}. Our implementation of \cref{algo:Sample} also removes \(v_{\src}\) from \(S\) before the return statement.

\subsection{Datasets}
We used real-world networks in our experiments. To ensure fairness in our experimental evaluation, we only selected networks that have already been used before for evaluating network embedding learning algorithms in the literature.
\begin{itemize}
    \item \textbf{BlogCatalog}~\cite{Tang:2009}: A network that depicts blogers' relationship. It has 10,312 nodes, 333,983 undirected edges\footnote{\label{ftn:incorrect-edge}This exact number of edges is different from that in~\cite{gl16}. We recounted this number and believe that \cite{gl16} might have made a mistake.} and 39 labels. The labels of the nodes represent the topic categories provided by authors.

    \item \textbf{Protein-Protein Interaction} (PPI)~\cite{gl16}: A network that depicts the interaction of proteins. We obtain labels from hallmark gene set~\cite{btr260}. The network has 3,890 nodes, 38,292 undirected edges\cref{ftn:incorrect-edge}, and 50 different labels.
    \item \textbf{Wikipedia}~\cite{gl16}: A co-occurrence network of words present in Wikipedia. It has 4,777 nodes, 92,091 undirected edges\cref{ftn:incorrect-edge}, and 40 different labels. 
\end{itemize}

\subsection{Node Classification}
\label{sec:node-classification}

Node classification is a classification task that, given the labels of a subset of nodes in \(G\), predicts the labels of the rest of the nodes in \(G\). It has been widely accepted as a standard instrument for evaluating network embedding learning algorithms~\cite{gf18}. 

To ensure fairness in our experimental evaluation, we deployed an experimental procedure that is commonly used in the literature. (a) First, for each network, we ran each network embedding learning algorithm once to learn a network embedding. (b) Then, we performed node classification using a \textit{one-versus-rest logistic regression classifier} (\texttt{scikit-learn}~\cite{sklearn} on top of \texttt{liblinear}~\cite{Fan2008} with 100 iterations): We randomly sampled a subset of nodes and used their representing vectors and labels as training data; and we used the representing vectors and labels of the rest of the nodes as test data. We repeated this step for a range of \textit{train ratios}, i.e., the ratios of the amount of training data to that of all data. For each train ratio, we repeated this step 10 times. (c) Finally, we evaluate predicted labels in the test data using the \textit{micro-f1} and \textit{macro-f1} scores averaged over the 10 runs of Step~(b) for each train ratio.

Due to the absence of dedicated multi-agent algorithms, we adapted \textit{Relational Neighbors} (RN)~\cite{macskassy2003simple}, a centralized algorithm, and used it as an additional baseline multi-agent algorithm. RN is an iterative algorithm for node classification. In each iteration, it traverses all nodes once and labels each node \(v\) with the most frequent label among its adjacent nodes \(\mathcal{N}(v)\). The algorithm terminates when all nodes have been labeled. RN can be easily adapted to be multi-agent: Each node \(v\) is maintained by an agent \(a(v)\). Then \(a(v)\) iteratively labels \(v\) with the most frequent label among \(\mathcal{N}(v)\). This only requires exchanging information locally between agents maintaining adjacent nodes and no information on the network would have been explicitly passed along via a chain of agents.

\begin{table}[t]
\centering
\caption{The optimal hyperparameters for MANELA and node2vec used in node classification.}\label{tab:opt-parameters}
\begin{tabular}{ccccc}
\toprule
     \textbf{Hyperparameters}& BlogCatalog & PPI & Wikipedia\\
     \midrule
     MANELA: \(r_1\) & 0.7 & 0.1 & 0.3\\
     node2vec: \(p,q\) & 0.25,0.25 & 4,1 & 4,0.5\\
     \bottomrule
\end{tabular}
\end{table}
 
In our experiments, we terminated each network embedding learning algorithm upon reaching the same number of times of update pairs. Following the recommended hyperparameters in~\cite{gl16}, we set \(w\) to 10, \(\ell\) to 80, and \(\gamma\) to 10 for both DeepWalk and node2vec. For MANELA, we also set \(w\) to 10 and made it perform the same number of times of update pairs as DeepWalk and node2vec. Following~\cite{pas14,gl16}, we set the dimension of learned vectors to 128. The learning rate of MANELA, DeepWalk, and node2vec is 0.025 initially and decreases to \(0.0001\times 0.025\) linearly with respect to the number of times that they performed update pairs. \Cref{tab:opt-parameters} lists other hyperparameters that we used. For node2vec, we used the optimized hyperparameters mentioned in~\cite{gl16}. Similarly, we ranged \(r_1\) from 0 to 1 and chose a value of \(r_1\) that led to the highest micro-f1 score for each network.

\begin{table}[t]
    \centering
    \caption{Micro-f1 scores in node classification resulted from each algorithm on each network where train ratios were set to 0.5.}\label{tab:micro-f1}
    \begin{subtable}{\linewidth}
    \centering
    \begin{tabular}{ccccc}
    \toprule
         \textbf{Algorithm}&\phantom{a}& BlogCatalog & PPI & Wikipedia  \\
         \midrule
         MANELA & &\(0.386\) & \(\mathbf{0.229}\) & \(0.492\) \\
         node2vec && \(\mathbf{0.394}\) & \(0.216\) &\(\mathbf{0.517}\) \\
         DeepWalk && \(0.380\) & \(0.210\) &\(0.474\) \\
         RN && \(0.037\) & \(0.039\) & \(0.005\) \\
         \bottomrule
    \end{tabular}
    \end{subtable}
\end{table}

\Cref{tab:micro-f1} summarizes our results. It lists the micro-f1 score resulted from each algorithm on each network with train ratios set to be 0.5. RN always had micro-f1 scores that are far lower than those of any other algorithms for all three networks.

For all three networks, MANELA had higher micro-f1 scores than those of DeepWalk. This might be caused by the fact that MANELA has a tunable capability of encapsulating homophily and structural equivalence. Mostly noticeably, in PPI, MANELA's micro-f1 score was as large as 9\% higher than DeepWalk.

MANELA had an overall performance that is comparable with that of node2vec: MANELA worked better on PPI, while node2vec worked better on BlogCatalog and Wikipedia. However, we emphasize that the settings of this comparison were biased against MANELA: (a) First, under our centralized experimental settings, while node2vec enjoyed unstinting exploitation of computational resources, our MANELA was unduly fettered by the AC. (b) Secondly, even if our experimental settings were made purely for evaluating centralized algorithms, they were still biased toward node2vec. According to our theoretical analysis in \cref{sec:node2vec-pre}, node2vec has a serious scalability issue that MANELA does not bear. To perform the same number of times of update pairs on the same network, node2vec demands much more computational resources, which can be onerous even for moderately large networks.

\begin{figure}[t]
    \centering
    \includegraphics[width=0.95\linewidth]{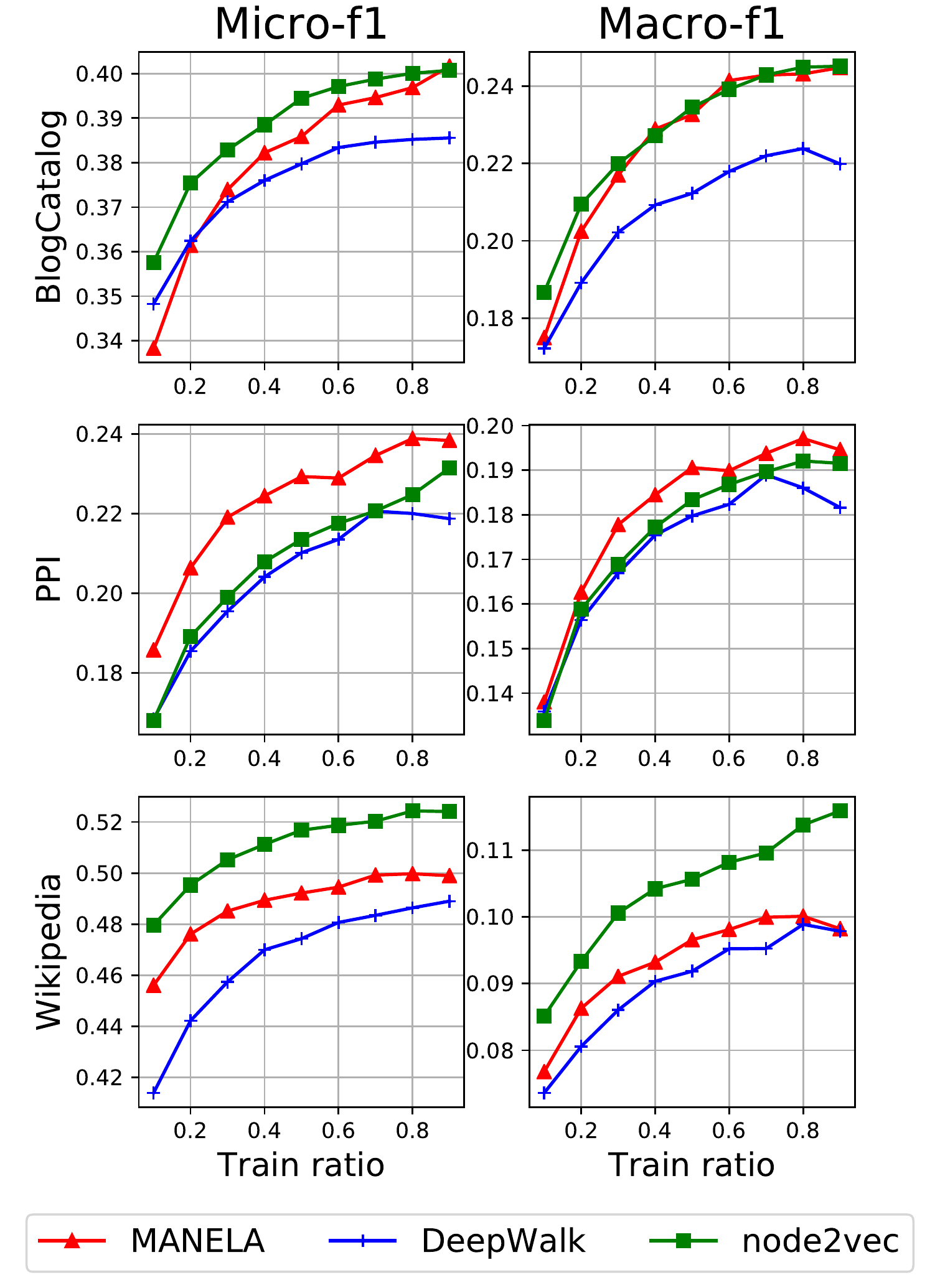}
    \caption{The results of node classification. Each row shows the results of a network. The y-axes are micro-f1 scores for the plots in the left column and are macro-f1 scores for those in the right column. The x-axes are train ratios, which range from 0.1 to 0.9 for all three networks.}\label{fig:node_classification}
\end{figure}

\Cref{fig:node_classification} shows more detailed comparisons among MANELA, DeepWalk, and node2vec. They further confirm our previous results: Although inured to the AC, MANELA was still more effective than DeepWalk; and it was also comparable to node2vec in terms of effectiveness, despite node2vec's undue demands on computational resources.

\subsection{Link prediction}
\begin{table}[t]
    \centering
    \caption{Averaged MAPs of 10 runs by all three algorithms. Similar to those in \cref{sec:node-classification}, we chose the values of the listed hyperparameters by searching for the ones that led to the highest averaged MAPs. We set all other hyperparameters to the same as in \cref{sec:node-classification} to achieve the same number of times that each algorithm performed update pairs.} \label{tab:map}
    \begin{tabular}{ccc}
        \toprule
        \textbf{Algorithm} & \textbf{Hyperparameters} & \textbf{Averaged MAP} \\\midrule
        MANELA  & \(r_1=0.4\) & 0.1360 \\
        node2vec & \(p=4,q=4\) & 0.1059 \\
        DeepWalk & None & 0.0964 \\
        \bottomrule
    \end{tabular}
\end{table}

\begin{figure}[t]
    \centering
    \includegraphics[width=0.8\linewidth]{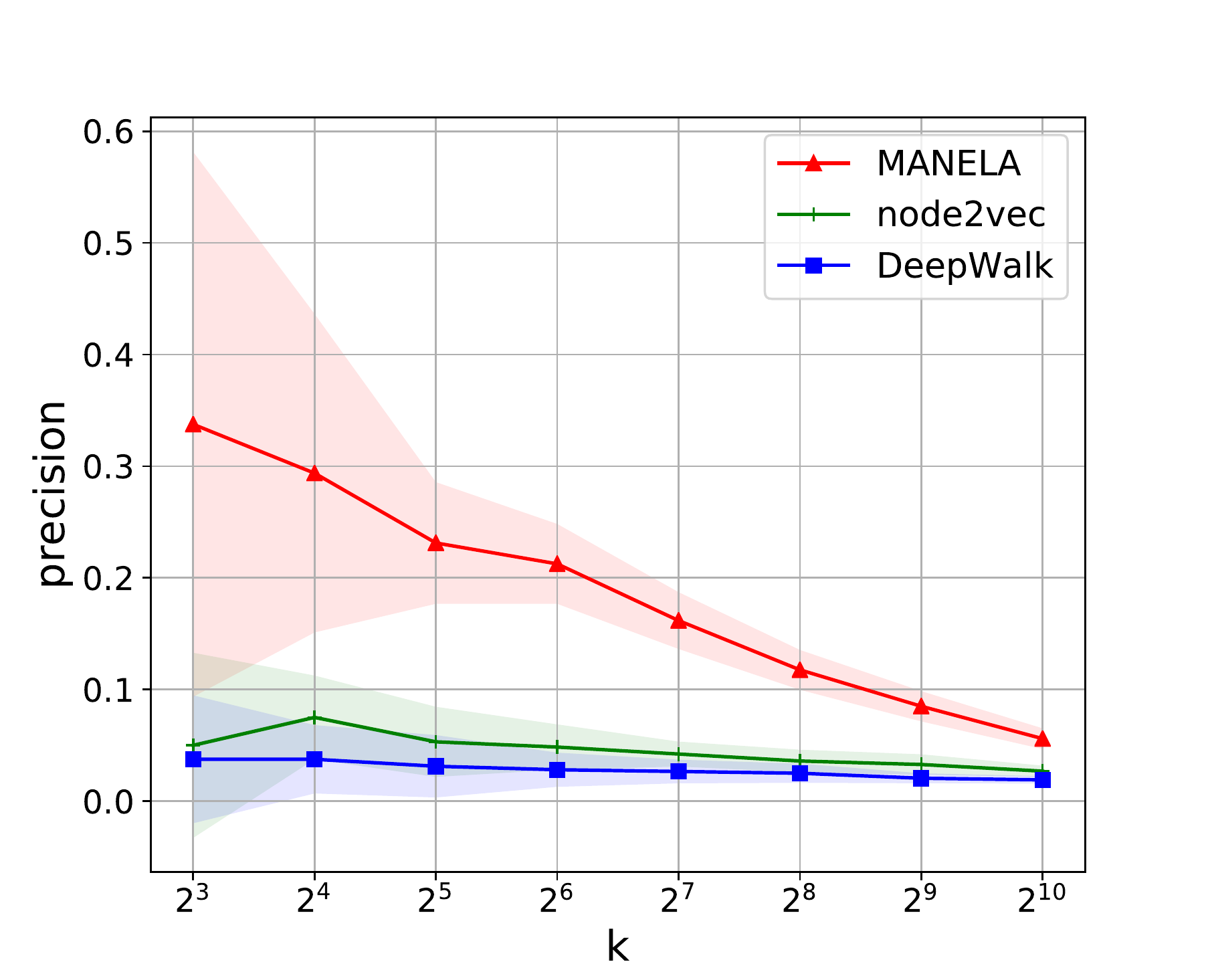}
    \caption{\(\preci k\) for different \(k\)'s averaged over 10 runs by MANELA, node2vec, and DeepWalk. Solid lines indicate the averages over \(\preci k\) produced in the 10 runs. Due to the heavy computational costs, the average \(\preci k\)'s were carried out on edges incident to a set \(V_{\text{s}}\) of 1024 randomly sampled nodes, i.e., we approximated \(E_{\text{pred}}\) in \cref{eq:prk} as \(E_{\text{pred}}\mid_{\bigcup_{v\in V_{\text{s}}} \partial v}\). The transparent area around the curves indicates the standard deviations for different \(k\)'s. }
    \label{fig:link}
\end{figure}
Link prediction is a rising application of network embeddings. A link prediction algorithm takes a network as input (i.e., training data), often in the context of a dynamically growing network, and predicts edges that would be added in the future. For example, in a social network, the algorithm predicts people (i.e., nodes) who are likely to get acquainted with each other in the future and thus helps recommend new connections to them. The predicted edges are usually in the form of a list (referred to as the \textit{predicted edge list} and denoted by \(E_{\text{pred}}\)) where predicted edges are sorted descendingly by their predicted likelihood to be added.

Since link prediction has not been popularly used to the evaluation of network embedding learning algorithms yet, we only experimented on PPI to show some preliminary results. In our experiment, we performed multiple runs and, in each run, we ran all three algorithms once. We generated the training and test data as follows. In each run, we randomly removed 20\% edges in PPI and used them as test data. In case the resulting subnetwork is disconnected, we further reduced it to the connected component with the maximum number of nodes. Then we used this further reduced subnetwork as training data for MANELA, DeepWalk, and node2vec to learn the network embeddings.

The score of a predicted edge \((v,v')\) is calculate using \(\Phi(v)\cdot \Phi(v')\). We use the \textit{precision at k}, denoted by \(\preci k\), and the \textit{mean average precision} (MAP) as the metrics for evaluating the effectiveness for link prediction. The higher they are, the more effective the algorithms are. Formally, \(\preci k\) is defined as
\begin{equation}\label{eq:prk}
    \preci k \coloneqq \frac{|E_{\text{pred}}[\{1,\ldots,k\}]\cap E_{\text{obs}}|}{k},
\end{equation}
where \(E_{\text{pred}}[\{1,\ldots,k\}]\) denotes the set of the first \(k\) edges in \(E_{\text{pred}}\) and \(E_{\text{obs}}\) is the test data (i.e., the set of the removed 20\% of the edges). The MAP is defined as
\begin{align}\label{eq:map}
E_{\text{obs}(v)} &\coloneqq E_{\text{obs}}\cap \partial v\\
E_{\text{pred}(v)} &\coloneqq E_{\text{pred}}\mid_{\partial v}\\
       \preci k(v) &\coloneqq \frac{\left |E_{\text{pred}(v)}[\{1,\ldots,k\}]\cap E_{\text{obs}(v)}\right |}{k}\\
       AP\left(v\right)&\coloneqq\frac{\sum_{k=1}^{|E_{\text{pred}(v)}|}\preci k\left(v\right)\cdot \mathbb{1}_{E_{\text{pred}(v)}[k]\in E_{\text{obs}(v)}}}{\max \left\{ \left |E_{\text{pred}(v)}\mid_{E_{\text{obs}(v)}}\right |, 1\right \}}\\
       MAP &\coloneqq \frac{\sum_{v\in V}AP\left(v\right)}{|V|},
\end{align}
where \(\partial v\) is the set of edges that are incident to \(v\); \(E_{\text{pred}}\mid_{\partial v}\) denotes the sublist of \(E_{\text{pred}}\) which only contains elements in both \(E_{\text{pred}}\) and \(\partial v\) and in which their ordering in \(E_{\text{pred}}\) is kept; and \(E_{\text{pred}(v)}[k]\) denotes the \(k^{\text{th}}\) element in  \(E_{\text{pred}(v)}\).

\cref{tab:map} summarizes our experimental results. MANELA clearly outperformed both node2vec and DeepWalk in terms of averaged MAPs. \Cref{fig:link} shows more detailed results with varying \(k\). These results further confirm the advantages of MENELA over node2vec and DeepWalk, especially when \(k\) was not too large.

\section{Discussion: Visualization}\label{sec:visualization}
\begin{figure}[t]
    \centering
    \begin{subfigure}[t]{0.32\linewidth}
        \centering
        \includegraphics[width=\textwidth]{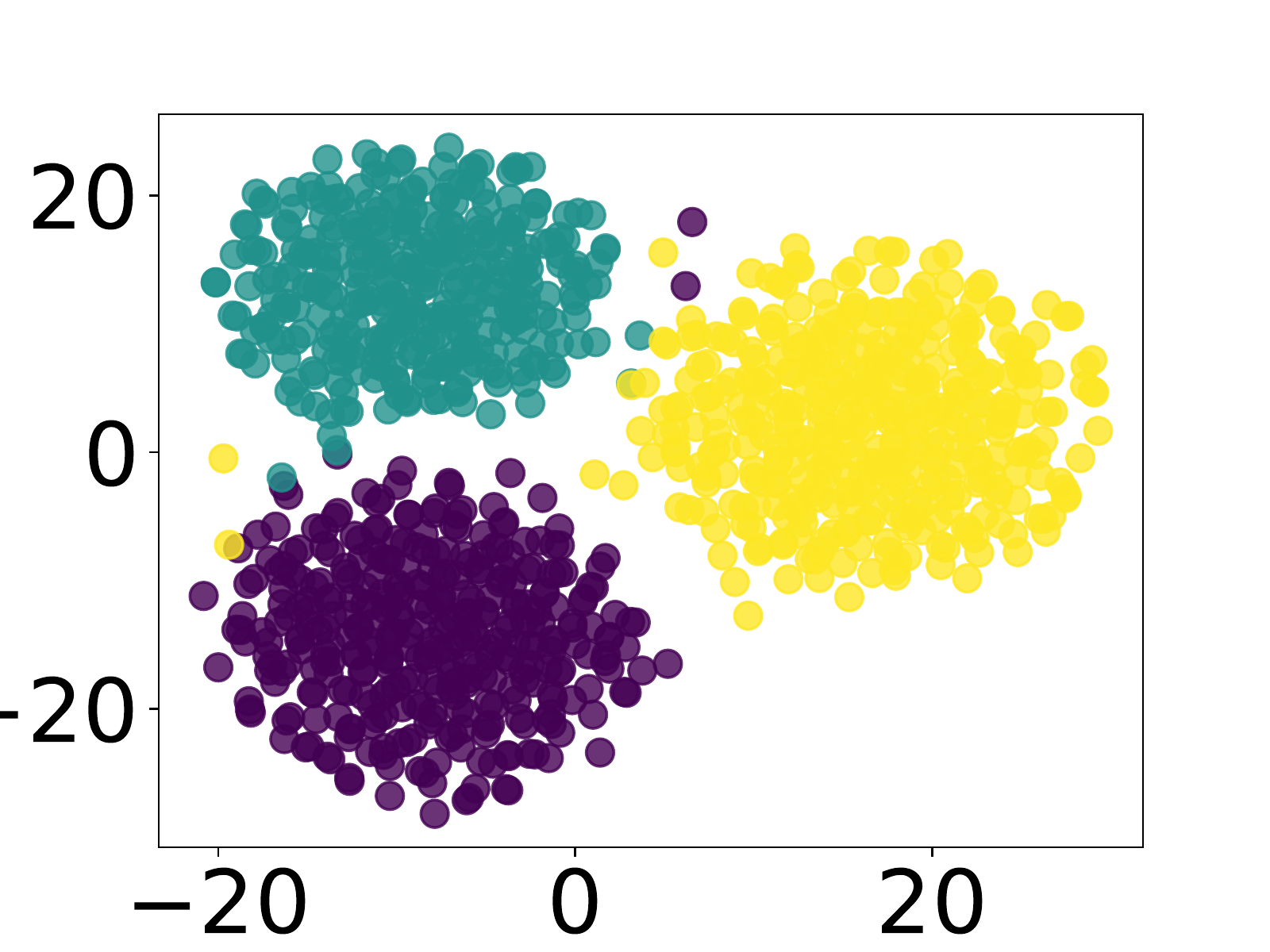}
        \caption{\(r_1=1\)}
    \end{subfigure}
    \begin{subfigure}[t]{0.32\linewidth}
        \centering
        \includegraphics[width=\textwidth]{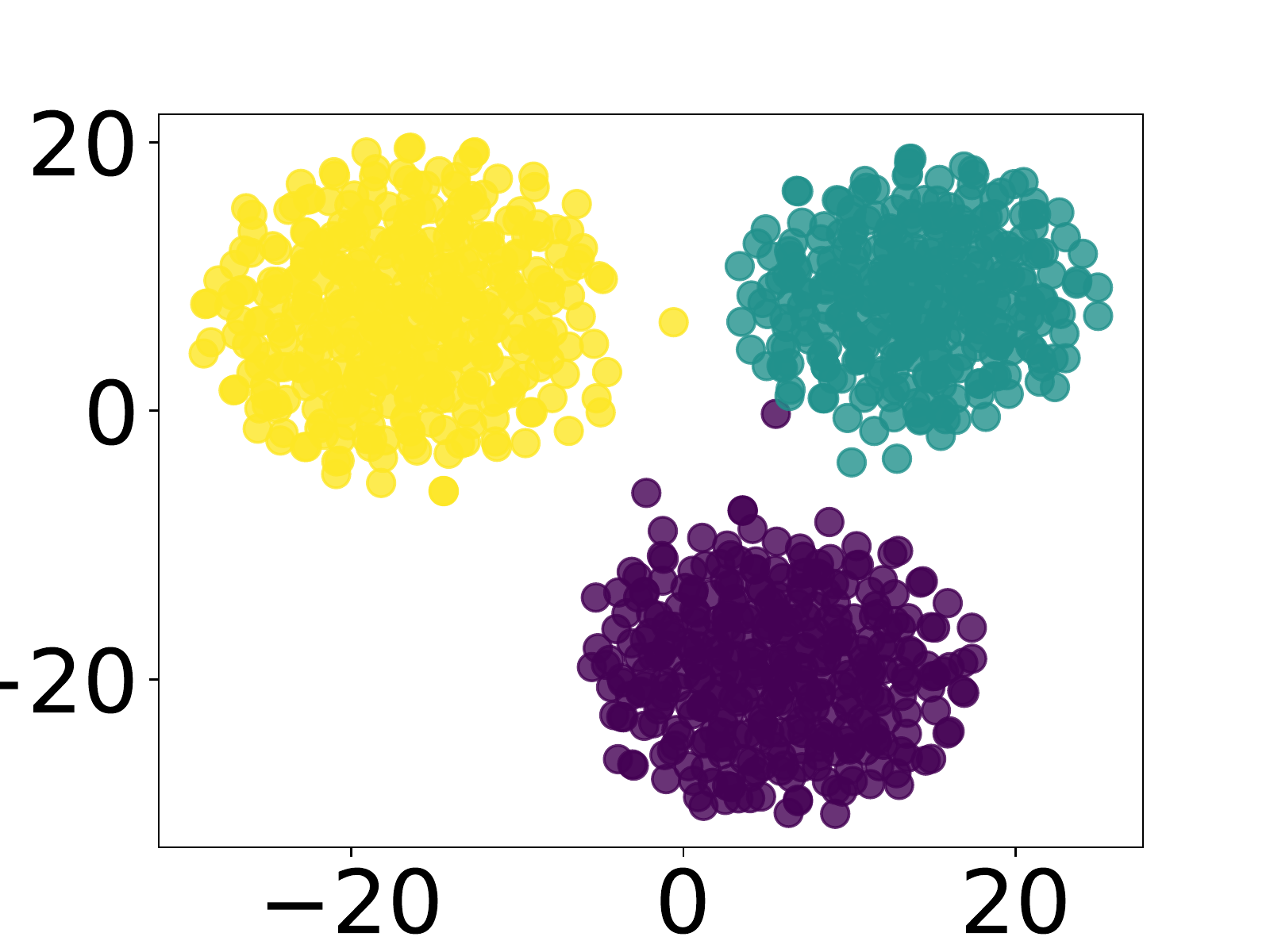}
        \caption{\(r_1=0.75\) }
    \end{subfigure}
    \begin{subfigure}[t]{0.32\linewidth}
        \centering
        \includegraphics[width=\textwidth]{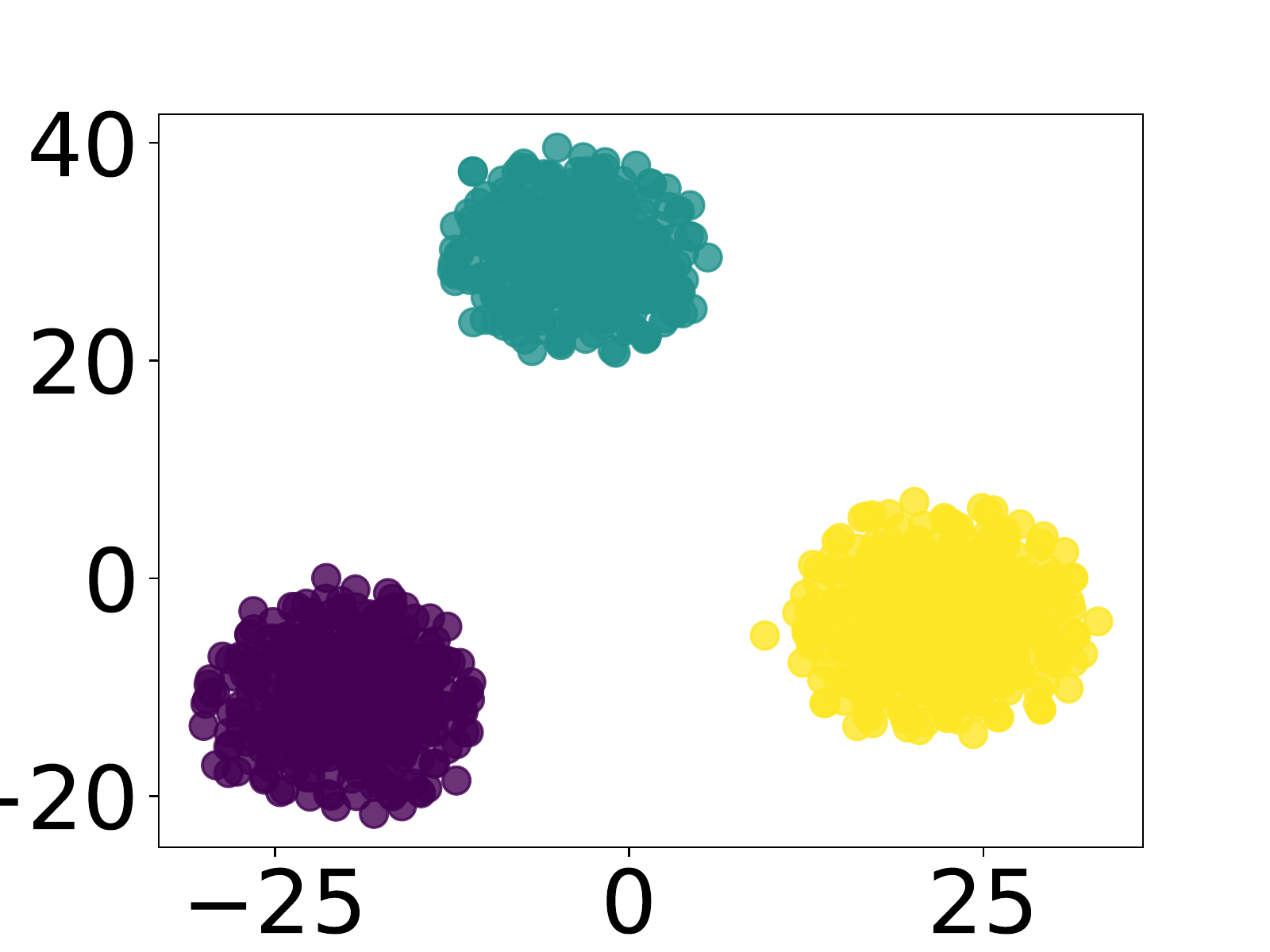}
        \caption{\(r_1=0.5\)}
    \end{subfigure}

    \caption{Visualization of the network embedding learned by MANELA on SYN-SBM. Each scattered dot represents the location of a node in the 2-D space. Every color indicates a different community.}
    \label{fig:visualization}
\end{figure}

Visualization is an important tool for understanding network embedding learning algorithms. To understand the network embeddings learned by MANELA, similar to \cite{gf18}, instead of using real-world networks, we used a random network, SYN-SBM~\cite{ycj1987}, which we deem more suitable for studying the encapsulation of homophily.

SYN-SBM is a random network generated using the \textit{stochastic block model} (SBM) with 1024 nodes, each of which belongs to a random community out of 3 communities. This network has 1024 nodes and 29,833 edges. The edges were added according to the following rules: For every two nodes in the same community, a edges had the probability of 0.1 to be added. Edges between nodes in different communities were added similarly except that the probability was 0.01. Considering homophily, nodes in the same community should tend to be more similar to each other than nodes in different communities. Therefore, ideally, nodes that belong to the same community were always clustered and there were clear boundaries between clusters corresponding to different communities.
    
We ran MANELA (with \(r_1=1, 0.75\), and 0.50) on SYN-SBM to learn network embeddings consisting of 128-D vectors. We used the \texttt{GEM} python package~\cite{gem2018} on top of the \textit{t-SNE}~\cite{mh2008} visualization tool to project learned 128-D vectors onto a 2-D vector space. The relative locations of nodes in the 2-D vector space approximately reflect their distances in the original 128-D vector space: Small distance between two nodes in the 2-D vector space usually implies a small distance in the original high-dimensional space as well. Therefore, the locations of these projected vectors provide an intuitive visualization of the similarity between nodes embodied in the network embeddings. 

\Cref{fig:visualization} plots these projected 2-D vectors. When \(r_1=1\), meaning that agents only performed update pairs with its adjacent nodes as the target nodes, the resulting boundaries between communities were somewhat blurred. However, when we gradually decreased \(r_1\) from 1 to 0.5, these boundaries were also gradually cleared up. This is because, when \(r_1\) decreased, the agents are more likely to sample nodes in its community other than its adjacent nodes. As a result, the sampled nodes, compared to those under the situation where \(r_1=1\), encoded a more accurate macro view of nodes' adjacency, which is critical in inferring homophily. Our intuitive understanding here is consistent with that in \cite{gl16}. As shown in this experiment, tuning \(\vec r\) effectively tuned the extent of encapsulating homophily.

\section{Discussion: Relationship to DeepWalk}
\label{sec:relation-to-deepwalk}

\begin{figure}[t]
    \centering
    \begin{subfigure}[b]{0.48\columnwidth}
        \centering
        \includegraphics[width=\linewidth]{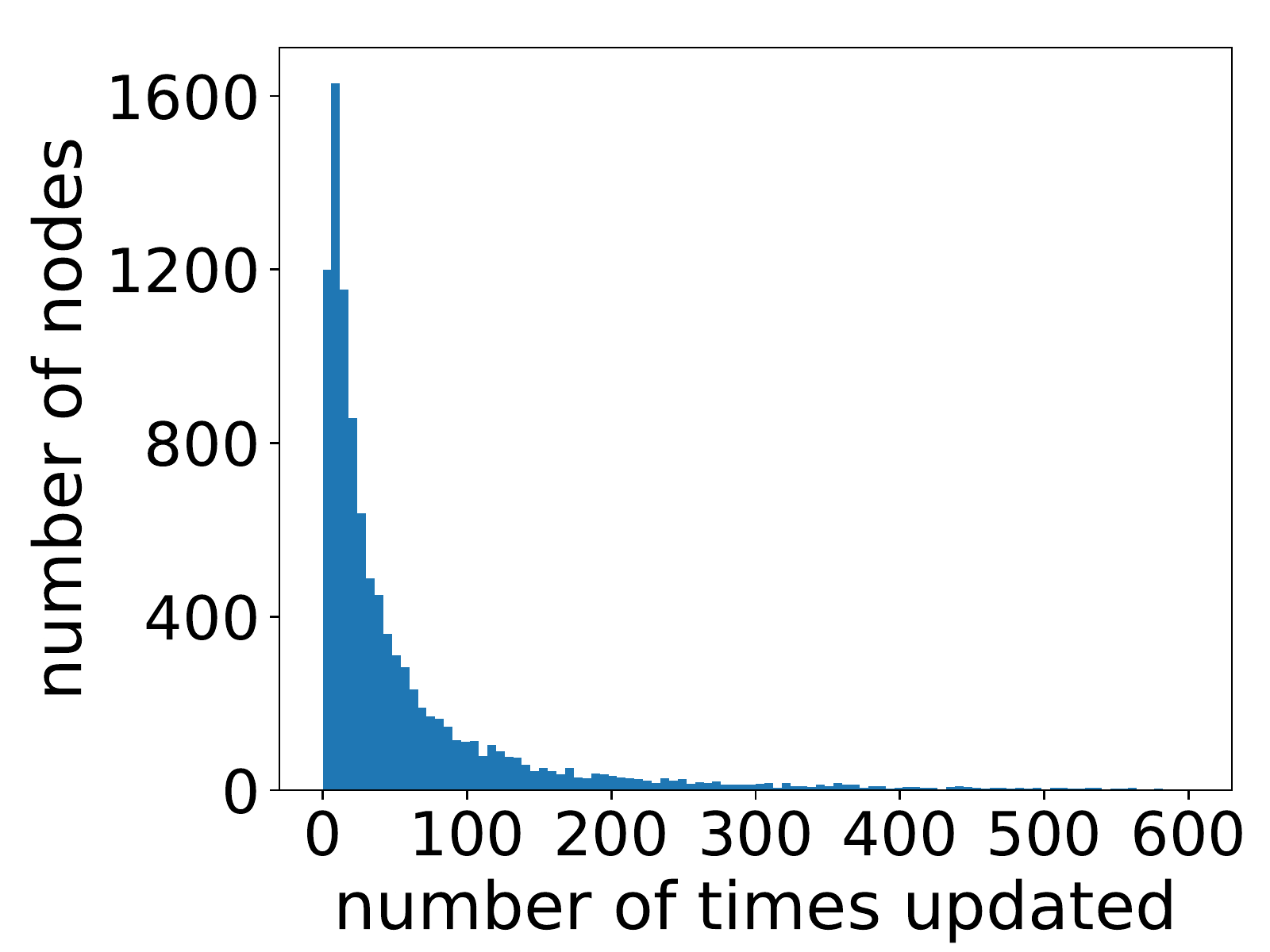}
        \caption{DeepWalk}
        \label{fig:hist_deepwalk}
    \end{subfigure}
    \hfill
    \begin{subfigure}[b]{0.48\columnwidth}
        \centering
        \includegraphics[width=\linewidth]{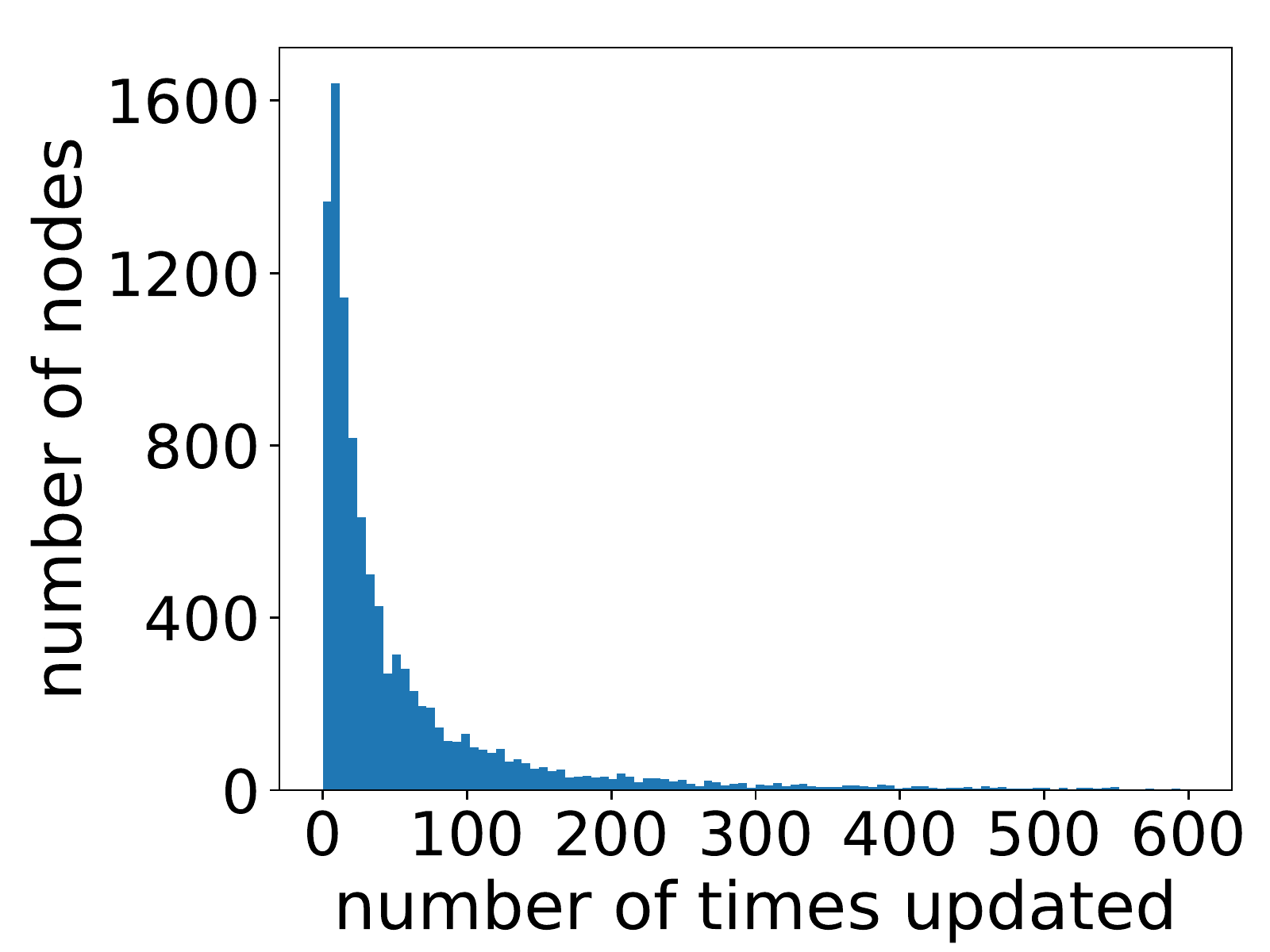}
        \caption{MANELA}
        \label{fig:hist_manela}
    \end{subfigure}

    \begin{subfigure}[b]{0.48\columnwidth}
        \centering
        \includegraphics[width=\linewidth]{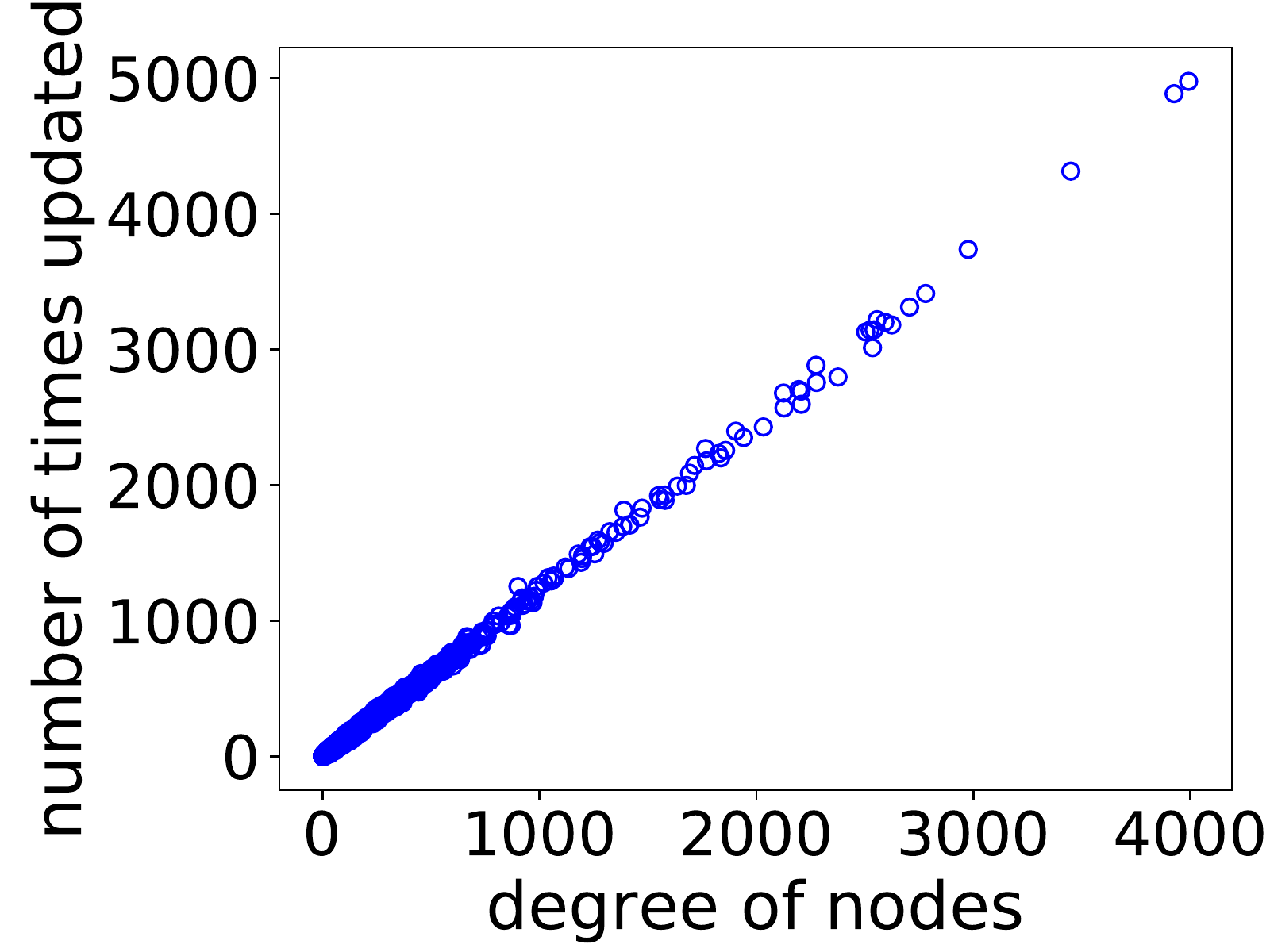}
        \caption{DeepWalk}
        \label{fig:scatter_deepwalk}
    \end{subfigure}
    \hfill
    \begin{subfigure}[b]{0.48\columnwidth}
        \centering
        \includegraphics[width=\linewidth]{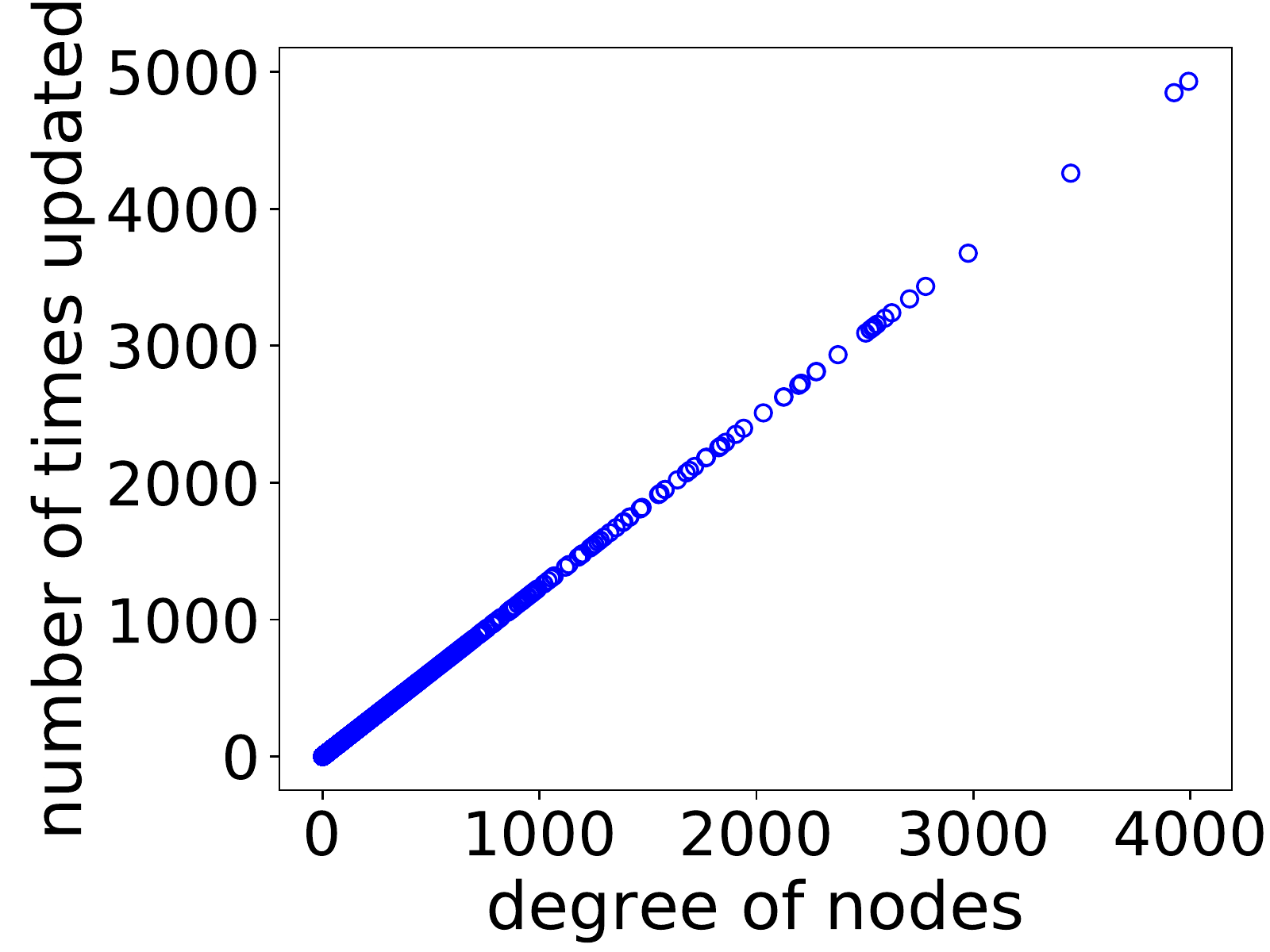}
        \caption{MANELA}
        \label{fig:scatter_manela}
    \end{subfigure}
    
    \caption{Similarity between Deepwalk and MANELA. The results are obtained during the node classification experiment on the BlogCatalog dataset in \cref{sec:experiment}. (a) and (b) show the histograms of the number of times that each node's representing vector was updated. Both follow power laws. (c) and (d) show the linear relationship between nodes' degrees (\(y\)-axes) and the number of times that their representing vectors were updated (\(x\)-axes). Each circle represents a node in the network.}
    \label{fig:similarity-manela-deepwalk}
\end{figure}

We further our understanding of MANELA by briefly exploring its relationship to DeepWalk. MANELA and DeepWalk are quite similar in many aspects. For example, both share very similar optimizing objectives and updating procedures on single vectors. In addition, both involve procedures that sample nodes for updating their representing vectors: DeepWalk employs random walks, while MANELA employs a multi-agent variant of it (\cref{algo:Sample} and \cref{alg:freq} in \cref{algo:agent}). Furthermore, as shown in \cref{fig:similarity-manela-deepwalk}, our experiment demonstrates the similarity between the effects of their sampling procedures with respect to the frequencies that nodes' representing vectors were updated. This similarity may partially explain why MANELA, while severely fettered by the AC, could still consistently achieve effectiveness that is comparable to (and in fact usually better than) that of DeepWalk.

\section{Conclusion}

In this paper, we proposed a multi-agent model for learning network embeddings. To the best of our knowledge, this is the first attempt to learn network embeddings under multi-agent settings. Under this model, we developed MANELA, a multi-agent algorithm for learning network embeddings. We demonstrated MANELA's advantages over DeepWalk and node2vec both theoretically and experimentally. Theoretically, MANELA is always the champion in terms of amortized time complexity, the capability of encapsulating different CSNs, and the multi-agentness. In our experimental evaluation, for node classification, MANELA was more effective than both DeepWalk and RN\@. Even if it was inured to the AC and conceded node2vec's undue demands on computational resources, MANELA was still able to achieve an effectiveness comparable to that of node2vec. For link prediction, we also preliminarily showed that MANELA beat both DeepWalk and node2vec on PPI\@. Finally, we furthered our understanding of MANELA via  visualization and exploration of its relationship to DeepWalk.
   
\bibliography{refs}
\bibliographystyle{IEEEtran}

\end{document}